\documentclass[twocolumn,showpacs,preprintnumbers]{revtex4}
\usepackage{graphicx}
\usepackage{dcolumn}
\usepackage{bm}
\newcommand{\ketbra}[1]{\ensuremath{| #1 \rangle \langle #1 |}}
\begin{document}

\title{Entanglement detection via condition of quantum correlation}
\author{Che-Ming Li$^{1,2}$}
\author{Li-Yi Hsu$^{3}$}
\author{Yueh-Nan Chen$^{4}$}
\author{Der-San Chuu$^{1}$}
\email{dschuu@mail.nctu.edu.tw}
\author{Tobias Brandes$^{5}$}
\affiliation{$^{1}$Department of Electrophysics, National Chiao Tung University, Hsinchu
30050, Taiwan}
\affiliation{$^2$Physikalisches Institut, Universit\"{a}t Heidelberg, Philosophenweg 12,
D-69120 Heidelberg, Germany}
\affiliation{$^{3}$Department of Physics, Chung Yuan Christian University, Chung-li
32023, Taiwan}
\affiliation{$^{4}$Department of Physics and National Center for Theoretical Sciences,
National Cheng Kung University, Tainan 701, Taiwan}
\affiliation{$^{5}$Institut f\"{u}r Theoretische Physik, Technische Universit\"{a}t
Berlin, Hardenbergstr. 36 D-10623 Berlin, Germany}
\date{\today }

\begin{abstract}
We develop a novel necessary condition of quantum correlation. It is
utilized to construct $d$-level bipartite Bell-type inequality which is
strongly resistant to noise and requires only analyses of $O(d)$ measurement
outcomes compared to the previous result $O(d^{2})$. Remarkably, a
connection between the arbitrary high-dimensional bipartite Bell-type inequality
and entanglement witnesses is found. Through the necessary condition of
quantum correlation, we propose that the witness operators to detect truly
multipartite entanglement for a generalized Greenberger-Horne-Zeilinger (GHZ)
state with two local measurement settings and a four-qubit singlet state with
three settings. Moreover, we also propose the first robust entanglement witness
to detect four-level tripartite GHZ state with only two local measurement
settings.
\end{abstract}

\maketitle

Entanglement is at the heart of quantum physics and a resource for quantum
information processing \cite{qip}. Multipartite entanglement for two-level
quantum systems (qubits) has attracted attention for its unusual features 
\cite{ghz} and necessity in a large-scale realization of quantum
computation and communication \cite{lcc}. In particular, with the rapid
development of technology for manipulating quantum states, multipartite
entanglement has been created experimentally and then utilized to quantum
information processing \cite{meqip}. In addition, entangled qubits, entanglement
for multi-level quantum systems (qudits) has been realized in few physical
systems \cite{qdexp}. Moreover, it has been proven that qudits have the
advantage over qubits \cite{improve}. Thus, identifying whether an
experiment's output is an entangled state for multipartite or multilevel
systems is very important for further studies on quantum correlation and to
perform reliable quantum protocols.

Bell-type inequalities (BIs) \cite{bell,chsh,collins} and entanglement
witnesses (EWs) \cite{witness, 1GMEEW,bouren,toth} are widely used to verify
quantum correlation. BIs are based on the local hidden variable theories, whereas EWs rely on an utilization of the whole or partial knowledge of the entangled state to be created. However, a single systematic approach to construct EWs for entangled qudits and to connect BIs for
arbitrary high-dimesional systems with EWs is still lacking. Investigations
on how entangled qudits can be shown efficiently and what is the fundamental
feature in entanglement verifications are both significant for a deeper
understanding of quantum correlation of qudits \cite{4x3} and for efficient
manipulations to achieve quantum information processing \cite{dc}.

In this work, we develop a novel necessary condition of quantum correlation.
This enables $d$-level bipartite BIs to be tested with only analyses of $O(d)$
measurement outcomes for detection events which is much smaller than the
previous result $O(d^{2})$ \cite{collins,fu}. In particular, a connection
between \textit{arbitrary high-dimensional} bipartite BIs and EWs is found. We
then use the correlator operators involved in the necessary condition of
quantum correlation to construct EWs for detecting \textit{genuine
multi-partite entanglement}, which can only be generated with participation
of \textit{all} parties of a system, about the generalized
Greenberger-Horne-Zeilinger (GHZ) state with two \textit{local measurement
settings} (LMSs) (which will be described in detail) and four-qubit singlet
states \cite{fourstate} with only three LMSs. More recently, it has been
shown that four qubit singlet state is very useful for quantum secret
sharing \cite{fourstate2}. Through our method, $15$ LMSs required for the EW
by Ref. \cite{bouren} can be reduced greatly. In order to show the high
generality of the condition of quantum correlation, we also give the first
EW to detect a four-level tripartite GHZ state \cite{4x3} with only two LMSs.
Moreover, the proposed EWs are resistant to noise. In what follows, an
introduction to the necessary condition of quantum correlation will be given
as a preliminary to further applications.

\section{Correlation conditions for quantum correlation}

In an experiment whose aim is to generate a multipartite entangled state $
\left|\xi\right\rangle$, if the experimental conditions are imperfect, it is important to know whether an experimental output state still possesses
multipartite quantum correlation wich is close to the state $
\left|\xi\right\rangle$. The first EW for detecting genuine multipartite
entanglement is given by Ref. \cite{1GMEEW} and formulated as: 
\begin{equation}
\mathcal{W}_{\xi}^{\text{p}}=\alpha_{\xi}^{\text{p}}\openone
-\left|\xi\right\rangle\left\langle\xi\right|,
\end{equation}
where $\alpha_{\xi}^{\text{p}}=\text{max}_{\left|\chi\right\rangle\in
B}|\left\langle\chi|\xi\right\rangle|^{2}$ and $B$ denotes the set of
biseparable states. Although it is difficult to determine the overlap $
\alpha_{\xi}^{\text{p}}$, through the general method proposed by Bourennane 
\textit{et al.} \cite{bouren}, one can perform this task. Thus, for some
experimental output state, say $\rho$, if measured outcomes show that $\text{
Tr}(\mathcal{W}_{\xi}^{\text{p}}\rho)<0$, the state $\rho$ is identified as
a genuine multipartite entanglement which is close to the state $
\left|\xi\right\rangle$.

It is worth noting that complete knowledge of the state $
\left|\xi\right\rangle$, i.e., all information about correlation characters,
is utilized for the witness operator, and, however, in order to measure the
operator $\mathcal{W}_{\xi}^{\text{p}}$ experimentally, the number of LMSs
appears to increase with the number of qubits of the state $
\left|\xi\right\rangle$ \cite{bouren}. A LMS, denoted by $M:(\hat{V}_{1},...,
\hat{V}_{n})$ in this paper, means that single-qubit measurements of
operator $\hat{V} _{i}$ for $i=1,...,n$ are taken on the $n$ remote parties
in parallel. In addition, EWs with forms such as $\mathcal{W}_{\xi}^{\text{p}}$,
the number of LMSs utilized to realize BIs typically increases exponentially
with the number of parties of the state. Moreover, the analyses of measured
outcomes for detection events also depend on the structures of BIs. A
detection event means a set of measurement outcomes, denoted by $
(v_{1},...,v_{n})$, under some LMS. For example, the LMS, $
M_{2z}=(\sigma_{z},\sigma_{z})$, corresponds to four possible detection
events: $(0,0)$, (0,1)$, (1,0)$, and $(1,1)$, where $v_{i}=0$ or $1$ stands
for the eigenvalue $(-1)^{v_{i}}$ of Pauli-operator $\sigma_{z}$. The
meaning of LMS and that of detection event are strictly different.

The witness operators proposed in this paper to detect genuine multipartite
entanglement have the following form: 
\begin{equation}
\mathcal{W}_{\xi}=\alpha_{\xi}\openone-\hat{C}_{\xi},
\end{equation}
where $\alpha_{\xi}$ is some constant and $\hat{C}_{\xi}$ is the operator
which is composed of several different kinds of correlator operators with \textit{necessary conditions of quantum correlations} imbedded in
the state $\left|\xi\right\rangle$. If outcomes of measurements show that $
\text{Tr}(\mathcal{W}_{\xi}\rho)<0$, the state $\rho$ is identified as a
truly multipartite entanglement. In what follows we will show that the
operator $\hat{C}_{\xi}$ can be constructed systematically and measured with
fewer LMSs for different kinds of pure multipartite entangled \textit{qubits}
or \textit{qudits}.

Furthermore, through the same idea behind the method to construct correlator
operators, a $d$-level bipartite BI is constructed and able to be tested
experimentally with fewer analyses of detection events. We then consider the
correlation conditions for quantum correlation involved in the approach to
construct correlator operators utilized in EWs and BIs as a connection
between them. We will see that the building blocks of the proposed EWs and
BIs are all derived from the correlation conditions for quantum correlation.

In order to present the idea behind the correlation condition for quantum
correlation clearly, let us first illustrate a derivation of correlation
condition for the generalized four-qubit GHZ state: 
\begin{equation}
\left|\Phi(\theta, \phi)\right\rangle=\cos (\theta
)\left|0000\right\rangle_{z}+e^{i\phi}\sin (\theta )\left|
1111\right\rangle_{z},
\end{equation}
for $0<\theta<\pi/4$ and $0\leq\phi<\pi/2$, where $
\left|v_{1}v_{2}v_{3}v_{4}\right\rangle_{z}=\otimes_{k=1}^{4}\left|v\right
\rangle_{kz}$ for $v\in\{0,1\}$ and $\left| v\right\rangle_{kz}$ corresponds
to an eigenstate of $\sigma _{z}$ with eigenvalue $(-1)^{v}$ for the party $k
$. For the four-qubit system, the kernel of our strategy for identifying
correlation between a specific subsystem, say $A$, and another one, say $B$,
under some LMS, $M_{l}$, relies on the sets of \textit{correlators} with the
following forms: 
\begin{eqnarray}
&&C_{0}^{(l)}=P(\text{v}_{A0},\text{v}_{B0})-P(\text{v}_{A1},\text{v}_{B0}),
\\
&&C_{1}^{(l)}=P(\text{v}_{A1},\text{v}_{B1})-P(\text{v}_{A0},\text{v}_{B1}),
\end{eqnarray}
where $P(\text{v}_{Ai},\text{v}_{Bj})$ is the joint probability for
obtaining the measured outcomes $\text{v}_{Ai}$ for the $A$ subsystem and $
\text{v}_{Bj}$ for the $B$ one. By the values of the correlators for an
experimental output state, we could identify correlations between outcomes
of measurements for the subsystems.

\textbf{Proposition 1.} If the results of measurements reveal that $C_{0}^{(l)}$
and $C_{1}^{(l)}$ are \textit{all positive} or \textit{all negative}, i.e., $
C_{0}^{(l)}C_{1}^{(l)}>0$, we are convinced that the outcomes of
measurements performed on the $A$ subsystem are correlated with the ones
performed on the $B$ subsystem.

\textit{Proof.} If the $A$ subsystem is independent of the $B$ one, we
recast $P(\text{v}_{Ai},\text{v}_{Bj})$ as $P(\text{v}_{Ai})P(\text{v}_{Bj})$
, where $P(\text{v}_{Ai})$ and $P(\text{v}_{Bj})$ denote the marginal
probabilities for obtaining results $\text{v}_{Ai}$ and $\text{v}_{Bj}$
respectively. Then, we have 
\begin{eqnarray}
&&C_{0,n}^{(l)}=(P(\text{v}_{A0})-P(\text{v}_{A1}))P(\text{v}_{B0}), \\
&&C_{1,n}^{(l)}=(P(\text{v}_{A1})-P(\text{v}_{A0}))P(\text{v}_{A1}).
\end{eqnarray}
Since $P(\text{v}_{A1}),P(\text{v}_{B0})\geq0$, we conclude that $
C_{0}^{(l)}C_{1}^{(l)}\leq 0$. Therefore, $C_{0}^{(l)}C_{1}^{(l)}> 0$
implies that the measured outcomes performed on the $A$ subsystem are
dependent with the one performed on the $B$ subsystem. Q.E.D.

We start showing the strategy with the help of proposition 1. Firstly, to
describe the correlation between \emph{a specific party} and \emph{others}
of the four-qubit system, we give four sets of correlator operators: 
\begin{eqnarray}
&&\hat{C}_{0,nz}^{(z)}=(\hat{0}_{nz}-\hat{1}_{nz})\otimes\hat{0}_{mz}\otimes
\hat{0}_{pz}\otimes\hat{0}_{qz}, \\
&&\hat{C}_{1,nz}^{(z)}=(\hat{1}_{nz}-\hat{0}_{nz})\otimes\hat{1}_{mz}\otimes
\hat{1}_{pz}\otimes\hat{1}_{qz},
\end{eqnarray}
for $n=1,...,4$, where $\hat{v}_{nz}= \left|v\right\rangle
_{nznz}\left\langle v\right|$ and $n$, $m $, $p$, and $q$ denote four
different parties under the LMS, $M_{4z}=(\sigma_{z},\sigma_{z},
\sigma_{z},\sigma_{z})$. In order to have compact forms, in what follows,
symbols of tensor product will be omitted from correlator operators. Then,
for some experimental output state, the expectation values of the hermitian
operators $\hat{C}_{0,n}^{(z)}$ and $\hat{C}_{1,n}^{(z)}$ are expressed in
the following correlators in terms of joint probabilities: 
\begin{eqnarray}
&&C_{0,n}^{(z)}=P(v_{n}=0,\text{v}=0)-P(v_{n}=1,\text{v}=0), \\
&&C_{1,n}^{(z)}=P(v_{n}=1,\text{v}=3)-P(v_{n}=0,\text{v}=3),
\end{eqnarray}
where $\text{v}= \sum_{i=1,i\neq n}^{4}v_{i}$. By proposition 1, we know
that if results of measurements reveal that $C_{0,n}^{(z)}C_{1,n}^{(z)}>0$,
we are convinced that the outcomes of measurements performed on the $n^{
\text{th}}$ party are correlated with the ones performed on the rest. If the 
$n^{\text{th}}$ party is independent of the rest, we have 
\begin{eqnarray}
&&C_{0,n}^{(z)}=(P(v_{n}=0)-P(v_{n}=1))P(\text{v}=0),  \nonumber \\
&&C_{1,n}^{(z)}=(P(v_{n}=1)-P(v_{n}=0))P(\text{v}=3),  \nonumber
\end{eqnarray}
and realize that $C_{0,n}^{(z)}C_{1,n}^{(z)}\leq 0$.

For the pure generalized four-qubit GHZ state, $\left|\Phi(\theta,\phi)
\right\rangle$, we have 
\begin{equation}
C_{0,n,\Phi(\theta,\phi)}^{(z)}=\cos^{2}(\theta),\
C_{1,n,\Phi(\theta,\phi)}^{(z)}=\sin^{2}(\theta),
\end{equation}
and hence $C_{0,n,\Phi(\theta,\phi)}^{(z)}C_{1,n,\Phi(\theta,\phi)}^{(z)}>0$
, which describes the outcomes of measurements are correlated. Then the
condition, $C_{0,n}^{(z)}C_{1,n}^{(z)}> 0$, is a \textit{necessary condition}
of the pure generalized four-qubit GHZ state.

Further, we construct the following correlator operators to identify
correlations between \emph{a specific group}, which is composed of the $n^{
\text{th}}$ and  $m^{\text{th}}$ parties, and \emph{another}: 
\begin{eqnarray}
&&\hat{C}_{0,nm}^{(z)}=(\hat{0}_{nz}\hat{0}_{mz}-\hat{1}_{nz}\hat{1}_{mz})
\hat{0}_{pz}\hat{0}_{qz}, \\
&&\hat{C}_{1,nm}^{(z)}=(\hat{1}_{nz}\hat{1}_{mz}-\hat{0}_{nz}\hat{0}_{mz})
\hat{1}_{pz}\hat{1}_{qz},
\end{eqnarray}
for $n,m=1,...,4$ and $n\neq m$. Moreover, we can express the expectation
values of the Hermitian operators $\hat{C}_{0,nm}^{(z)}$ and $\hat{C}
_{1,nm}^{(z)}$ in terms of joint probabilities for some output state: 
\begin{eqnarray}
&&C_{0,nm}^{(z)}=P(v_{nm}=0,\text{v}^{\prime }=0)-P(v_{nm}=2,\text{v}
^{\prime }=0),\ \ \ \ \ \  \\
&&C_{1,nm}^{(z)}=P(v_{nm}=2,\text{v}^{\prime }=2)-P(v_{nm}=0,\text{v}
^{\prime }=2),\ \ \ \ \ \ 
\end{eqnarray}
where $v_{nm}=v_{n}+v_{m}$ and $\text{ v}^{\prime }=\sum_{i=1,i\neq n\neq
m}^{4}v_{i}$. Proposition 1 shows that if the subsystem composed of the $n^{
\text{th}}$ and the $m^{\text{th}}$ parties is uncorrelated with another
one, the measured outcomes must satisfy $C_{0,nm}^{(z)}C_{1,nm}^{(z)}\leq 0$
. On the other hand, $C_{0,nm}^{(z)}C_{1,nm}^{(z)}>0$ indicates that they
are dependent.

It is clear that, for a pure generalized four-qubit GHZ state, we have 
\begin{equation}
C_{0,nm,\Phi (\theta ,\phi )}^{(z)}=\cos ^{2}(\theta ),\ C_{1,nm,\Phi
(\theta ,\phi )}^{(z)}=\sin ^{2}(\theta ),
\end{equation}
and hence $C_{0,nm,\Phi (\theta ,\phi )}^{(z)}C_{1,nm,\Phi (\theta ,\phi
)}^{(z)}>0$. Thus we know that the subsystem composed of the $n^{\text{th}}$
and the $m^{\text{th}}$ parties are correlated with another. Therefore, the
condition, $C_{0,nm}^{(z)}C_{1,nm}^{(z)}>0$, is also a necessary condition
of the state $\left\vert \Phi (\theta ,\phi )\right\rangle $.

After introducing two correlation conditions for the pure generalized GHZ
state under $M_{4z}$, let us progress towards the third one for correlation.
Under the LMS, $M_{4x}=( \sigma_{x},\sigma_{x},\sigma_{x},\sigma_{x})$, we
formulate four sets of correlators which correspond to the following
operators for identifying correlations between the $n^{\text{th}}$ party and
others: 
\begin{eqnarray}
&&\hat{C}_{0,n}^{(x)}=(\hat{0}_{nx}-\hat{1}_{nx})\otimes\hat{\mathbf{E}}, \\
&&\hat{C}_{1,n}^{(x)}=(\hat{1}_{nx}-\hat{0}_{nx})\otimes\hat{\mathbf{O}},
\end{eqnarray}
where 
\begin{eqnarray}
&&\hat{\mathbf{E}}=(\hat{0}_{mx}\hat{0}_{px}\hat{0}_{qx}+ \hat{0}_{mx}\hat{1}
_{px}\hat{1}_{qx}  \nonumber \\
&&\ \ \ \ \ \ \ +\hat{1}_{mx}\hat{0}_{px}\hat{1}_{qx}+\hat{1} _{mx}\hat{1}
_{px}\hat{0}_{qx}), \\
&&\hat{\mathbf{O}}=(\hat{1}_{mx}\hat{1}_{px}\hat{1}_{qx}+\hat{1}_{mx}\hat{0}
_{px}\hat{0}_{qx}  \nonumber \\
&&\ \ \ \ \ \ \ +\hat{0} _{mx}\hat{1}_{px}\hat{0}_{qx}+\hat{0}_{mx}\hat{0}
_{px}\hat{1}_{qx}).
\end{eqnarray}
From the expectation values of $\hat{C}_{0,n}^{(x)}$ and $\hat{C}_{1,n}^{(x)}
$ for some state and proposition 1, we could know the correlation behavior
of the system, i.e., for a system in which the $n^{\text{th}}$ party is
uncorrelated with the rest under $M_{4x}$, the outcomes of measurements must
satisfy the condition: $C_{0,n}^{(x)}C_{1,n}^{(x)}\leq 0$.

For the pure state, $\left|\Phi(\theta,\phi)\right\rangle$, the expectation
values of $\hat{C}_{k,n}^{(x)}$ is given by 
\begin{equation}
C_{0,n,\Phi(\theta,\phi)}^{(x)}=C_{1,n,\Phi(\theta,\phi)}^{(x)}=\sin
(2\theta)\cos(\phi)/2,
\end{equation}
and ensure that there are correlations between measured outcomes under the
LMS, $M_{4x}$. Thus the condition, $C_{0,n}^{(x)}C_{1,n}^{(x)}>0$, is
necessary for the pure generalized four-qubit GHZ state.

Entanglement imbedded in the pure generalized four-qubit GHZ state manifests
itself via necessary conditions of correlations presented above under two
LMSs. Therefore we combine all of the correlator operators involved in the
necessary conditions: 
\[
\hat{C}_{\Phi }=\hat{C}^{(z)}+\hat{C}^{(x)},
\]
where 
\begin{eqnarray}
\hat{C}^{(z)} &=&\sum_{j=0}^{1}(\sum_{n=1}^{4}\hat{C}_{j,n}^{(z)}+
\sum_{m=2}^{4}\hat{C}_{j,1m}^{(z)})  \nonumber \\
&=&8(\hat{0}_{1z}\hat{0}_{2z}\hat{0}_{3z}\hat{0}_{4z}+\hat{1}_{1z}\hat{1}
_{2z}\hat{1}_{3z}\hat{1}_{4z})-\openone, \\
\hat{C}^{(x)} &=&\sum_{n=1}^{4}\sum_{k=0}^{1}\hat{C}_{k,n}^{(x)}=4\sigma
_{x}\sigma _{x}\sigma _{x}\sigma _{x},
\end{eqnarray}
and $\openone$ is an identify operator, and then utilize the operator $\hat{C
}_{\Phi }$ to construct witness operator for detections of truly
multipartite entanglement. Three example are shown as follows. The witness
operator: 
\begin{equation}
\mathcal{W}_{\Phi }(\theta ,\phi )=\alpha _{\Phi }(\theta ,\phi )\openone-
\hat{C}_{\Phi },
\end{equation}
where $\alpha _{\Phi }(\theta ,\phi )$ is some constant, detects genuine
multipartite entanglement for the cases, $(\theta ,\phi )$: $(\pi /4,\pi /6)$
, $(\pi /4.9,0)$, and $(\pi /3.7,\pi /9)$. TABLE I gives a summary of $
\alpha _{\Phi }(\theta ,\phi )$ for these cases.

In order to prove that $\mathcal{W}_{\Phi}(\theta,\phi)$ is a EW for
detecting genuine multipartite entanglement, we have to show the following
comparison between 
\begin{equation}
\mathcal{W}_{\Phi}^{\text{p}}(\theta,\phi)=\alpha_{\Phi}^{\text{p}}\openone
-\left|\Phi(\theta,\phi)\right\rangle\left\langle\Phi(\theta,\phi)\right|,
\end{equation}
and $\mathcal{W}_{\Phi}(\theta,\phi)$ \cite{toth}: if a state $\rho$
satisfies $\text{Tr}(\mathcal{W}_{\Phi}(\theta,\phi)\rho)<0$, it also
satisfies $\text{Tr}(\mathcal{W}_{\Phi}^{\text{p}}(\theta,\phi)\rho)<0$,
i.e., $\mathcal{W}_{\Phi}(\theta,\phi)-\gamma_{\Phi}\mathcal{W}_{\Phi}^{
\text{p}}(\theta,\phi)\geq0$, where $\gamma_{\Phi}(\theta,\phi)$ is some
positive constant. Through the method given by Bourennane \textit{et al.} 
\cite{bouren}, we derive the operator $\mathcal{W}_{\Phi}^{\text{p}
}(\theta,\phi)$ and have $\alpha_{\Phi}^{\text{p}}=\cos^{2}(\theta)$ for $
0<\theta\leq\pi/4$ and $\alpha_{\Phi}^{\text{p}}=\sin^{2}(\theta)$ for $
\pi/4\leq\theta<\pi/2$. Table I summarizes the parameters $\gamma_{\Phi}$
utilized to prove that the proposed operators are indeed EWs for detecting
truly multipartite entanglement.

\begin{table}[tbp]
\caption{Summaries of numerical results of $\protect\alpha_{\Phi}(\protect
\theta,\protect\phi)$ for $\mathcal{W}_{\Phi}(\protect\theta,\protect\phi)$,
the parameters, $\protect\gamma_{\Phi}$, which are utilized to prove $
\mathcal{W}_{\Phi}(\protect\theta,\protect\phi)$ and $\protect\delta_{\text{
noise},\Phi}$ involved in robustness of the proposed witness operator for
detecting truly multipartite entanglement. Three different cases for the
state $\left|\Phi(\protect\theta,\protect\phi)\right\rangle$ corresponding
to $\mathcal{W}_{\Phi}(\protect\theta,\protect\phi)$ have been demonstrated.
\newline
}
\label{tab:table1}
\begin{ruledtabular}
\begin{tabular}{cccc}
$(\theta,\phi)$ & $(\frac{\pi}{4},\frac{\pi}{6})$ & $(\frac{\pi}{4.9},0)$ & $(\frac{\pi}{3.7},\frac{\pi}{9})$ \\
\hline
$\alpha_{\Phi}$ & $9.01$ & $9.21$ & $8.92$ \\
$\gamma_{\Phi}$ & $6.54$ & $6.44$ & $6.86$ \\
$\delta_{\text{noise},\Phi}$ & $0.139$ &  $0.150$ &  $0.169$ \\
\end{tabular}
\end{ruledtabular}
\end{table}

In addition, we are concerned with the robustness to noise for the witness $
\mathcal{W}_{\Phi}(\theta,\phi)$. The robustness of $\mathcal{W}
_{\Phi}(\theta,\phi)$ depends on the noise tolerance: $p_{\text{noise}
}<\delta_{\text{noise}}$, is such that 
\begin{equation}
\rho=\frac{p_{\text{noise}}}{2^{N}}\openone+(1-p_{\text{noise}
})\left|\Phi(\theta,\phi)\right\rangle\left\langle\Phi(\theta,\phi)\right|,
\end{equation}
where $p_{\text{noise}}$ describes the noise fraction, is identified as a
genuine multipartite entanglement. Three cases for the robustness to noise
for the witness $\mathcal{W}_{\Phi}(\theta,\phi)$ have been summarized in
Table I.

Further, we show the expectation values of the proposed EWs for different
pure states by Table II. From comparison with the results we know that a aim
state, say $\left|\Phi(\theta^{\prime },\phi^{\prime })\right\rangle$, does
not always give the smallest expectation value of the corresponding witness
operator, $\mathcal{W}_{\Phi}(\theta^{\prime },\phi^{\prime })$. One can
identify with the operator $\mathcal{W}_{\Phi}(\theta^{\prime },\phi^{\prime
})$ that an experimental output $\rho$ is truly multipartite entanglement if 
$\text{Tr}(\mathcal{W}_{\Phi}(\theta^{\prime },\phi^{\prime })\rho)<0$.
Further, if $\text{Tr}(\mathcal{W}_{\Phi}(\theta^{\prime },\phi^{\prime
})\rho)<\text{Tr}(\mathcal{W}_{\Phi}(\theta^{\prime },\phi^{\prime })|
\Phi(\theta^{\prime },\phi^{\prime }) \rangle \langle \Phi(\theta^{\prime
},\phi^{\prime }) | )$, the state $\rho$ is not in the state $
\left|\Phi(\theta^{\prime },\phi^{\prime })\right\rangle$ class.

The novel approach to derive $\hat{C}_{\Phi }$ shown above can be applied to
the cases for arbitrary number of qubits straightforwardly. One can
formulate sets of correlator operators to identify correlations between two
subsystems under two LMSs and then construct the witness operators further.
In particular, we have found that the proposed method also provides an
analytical and systematic way to construct correlator operators for
entangled states with local stabilizers and the corresponding EWs as the
previous results \cite{toth,further}.

Before proceeding further, let us give a brief summary and conclusion for
this section. We have demonstrated a systematical method to derive
correlator operators utilized to construct witness operators. The proposed
correlator operators are based on novel necessary conditions of some pure
multipartite entangled state to be created experimentally. Moreover, in the
example, these witness operators can be measured with only two LMSs. In what
follows, we will give two novel EWs in which the correlator operators can be
constructed systematically. Through these cases for entanglement detection,
one could realize that the proposed conditions of quantum correlations
possess a wide generality.

\begin{table}[tbp]
\caption{Expectation values of three proposed EWs including $\mathcal{W}
_{\Phi}(\frac{\protect\pi}{4},\frac{\protect\pi}{6})$, $\mathcal{W}_{\Phi}(
\frac{\protect\pi}{4.9},0)$, and $\mathcal{W}_{\Phi}(\frac{\protect\pi}{3.7},
\frac{\protect\pi}{9})$ for the pure states $\left|\Phi\right\rangle$: $
\left|\Phi(\frac{\protect\pi}{4},\frac{\protect\pi}{6})\right\rangle$, $
\left|\Phi(\frac{\protect\pi}{4.9},0)\right\rangle$, and $\left|\Phi(\frac{
\protect\pi}{3.7},\frac{\protect\pi}{9})\right\rangle$.}
\label{tab:table2}
\begin{ruledtabular}
\begin{tabular}{cccc}
 $\left|\Phi\right\rangle$& $\left|\Phi(\frac{\pi}{4},\frac{\pi}{6})\right\rangle$ & $\left|\Phi(\frac{\pi}{4.9},0)\right\rangle$ & $\left|\Phi(\frac{\pi}{3.7},\frac{\pi}{9})\right\rangle$ \\
\hline
$\text{Tr}(\mathcal{W}_{\Phi}(\frac{\pi}{4},\frac{\pi}{6})\ketbra{\Phi})$ & $-1.45$ & $-1.83$ & $-1.72$ \\
$\text{Tr}(\mathcal{W}_{\Phi}(\frac{\pi}{4.9},0)\ketbra{\Phi})$ & $-1.25$ & $-1.63$ & $-1.52$ \\
$\text{Tr}(\mathcal{W}_{\Phi}(\frac{\pi}{3.7},\frac{\pi}{9})\ketbra{\Phi})$ & $-1.55$ &  $-1.92$ &  $-1.81$ \\
\end{tabular}
\end{ruledtabular}
\end{table}

\section{EWs for multipartite entangled states}

\subsection{Detection of genuine multipartite entanglement of the four-qubit
singlet state}

Very recently, four-party quantum secret sharing has been demonstrated via
the resource of four photon entanglement \cite{fourstate2}, which is called
the four-qubit singlet state \cite{fourstate}. Through the same method
presented in the Introduction, we give a novel EW to detect the four-qubit
singlet state.

The four-qubit singlet state is expressed as the following form: 
\begin{eqnarray}
\left|\Psi\right\rangle =&&\frac{1}{\sqrt{3}}(\left| 0011\right\rangle_{z}
+\left| 1100\right\rangle_{z}  \nonumber \\
&&-\frac{1}{2}(\left| 0110\right\rangle _{z}+\left| 1001\right\rangle_{z}
+\left| 0101\right\rangle_{z} +\left| 1010\right\rangle_{z} )).  \nonumber \\
\end{eqnarray}
Under the LMS, $M_{4z}$, we formulate eight sets of criteria for identifying
quantum correlation between a specific party and others: the first type of 
identifications include the following four sets of correlators: 
\begin{eqnarray}
&&\hat{C}_{0,m}^{(z)}=\hat{0}_{1z}\hat{0}_{2z}\hat{1}_{3z}\hat{1}_{4z}-X_{m}(
\hat{0}_{1z}\hat{0}_{2z}\hat{1}_{3z}\hat{1}_{4z})X_{m}, \\
&&\hat{C}_{1,m}^{(z)}=\hat{1}_{1z}\hat{1}_{2z}\hat{0}_{3z}\hat{0}_{4z}-X_{m}(
\hat{1 }_{1z}\hat{1}_{2z}\hat{0}_{3z}\hat{0}_{4z})X_{m},
\end{eqnarray}
where $X_{m}=\sigma_{x}$ is performed on the $m^{\text{th}}$ party for $
m=1,...,4$. Then, the second type criteria are formulated as: 
\begin{eqnarray}
&&\hat{C} _{0n,k}^{(z)}=(\hat{0}_{(2n+1)z}\hat{1}_{(2n+2)z}-X_{k}(\hat{0}
_{(2n+1)z} \hat{1} _{(2n+2)z})X_{k})  \nonumber \\
&&\ \ \ \ \ \ \ \ \ \ \ \ (\hat{0}_{(2n\oplus 3)z}\hat{1}_{(2n\oplus 4)z}+
\hat{1} _{(2n\oplus 3)z} \hat{0}_{(2n\oplus 4)z}), \\
&&\hat{C}_{1n,k}^{(z)}=(\hat{1}_{(2n+1)z}\hat{0} _{(2n+2)z}-X_{k}(\hat{1}
_{(2n+1)z} \hat{0}_{(2n+2)z})X_{k})  \nonumber \\
&&\ \ \ \ \ \ \ \ \ \ \ \ (\hat{0}_{(2n\oplus 3)z}\hat{1}_{(2n\oplus 4)z}+
\hat{1} _{(2n\oplus 3)z}\hat{0}_{(2n\oplus4)z}),
\end{eqnarray}
where $k=(2n+1),(2n+2)$ for $n=0,1$; and the symbol "$\oplus$" behaves as
the addition of modulo $4$ for $n=1$ and as an ordinary addition for $n=0$.
The expectation values of the operators $\hat{C}_{l,m}^{(z)}$ and $\hat{C}
_{ln,k}^{(z)}$ for the pure four-qubit singlet state can be evaluated
directly and are given by $C_{l,m,\Psi}^{(z)}=1/3$ and $C_{ln,k,
\Psi}^{(z)}=1/6$ for $l=0,1$.

It is easy to see that the conditions involved in the expectation values of $
\hat{C} _{l,m}^{(z)}$ and $\hat{C} _{ln,k}^{(z)}$: 
\begin{eqnarray}
&&C_{0,m}^{(z)}C_{1,m}^{(z)}>0\text{ and }C_{0n,k}^{(z)}C_{1n,k}^{(z)}>0,
\end{eqnarray}
are necessary for the pure four-qubit singlet state. The proof of this
statement is similar to the one for proposition 1 presented in the first
section.

For invariance of the wave function presented in the eigenbasis of $\sigma
_{x}$ ($\sigma _{y}$), in analogy, we can construct eight sets of Hermitian
operators, 
\begin{eqnarray}
(\hat{C}_{0,m}^{(x(y))},\hat{C} _{1,m}^{(x(y))}) \text{ and } (\hat{C}
_{0n,k}^{(x(y))},\hat{C}_{1n,k}^{(x(y))}),  \nonumber
\end{eqnarray}
via the replacement of the index $z$ in above hermitian operators by the
index $x$ ($y$) and constructing the operators in the eigenbasis of $\sigma
_{x(y)} $. The expectation values of the above operators are all positive
for the state $\left|\Psi\right\rangle$, and so we have the following
necessary conditions of the state $\left|\Psi\right\rangle$: 
\begin{eqnarray}
&&C_{0,m}^{x(y)}C_{1,m}^{(x(y))}>0\text{ and }
C_{0n,k}^{(x(y))}C_{1n,k}^{(x(y))}>0,
\end{eqnarray}
Then, we combine all of the correlator operators proposed above:  
\begin{equation}
\hat{C}_{\Psi}=\hat{C}^{(x)}_{\Psi}+\hat{C}^{(y)}_{\Psi}+\hat{C}
^{(z)}_{\Psi},
\end{equation}
where 
\begin{equation}
\hat{C}^{(i)}_{\Psi}=\sum_{l=0}^{1}\big( 5\sum_{m=1}^{4}\hat{C}
_{l,m}^{(i)}+\sum_{n=0}^{1}\sum_{k=2n+1}^{2n+2}\hat{C} _{ln,k}^{(i)}\big),
\end{equation}
for $i=x,y,z$, and present a EW to detect the four-qubit singlet state. The
following witness operator detects truly multipartite entanglement for
states close to the state $\left|\Psi\right\rangle$: 
\begin{equation}
\mathcal{W}_{\Psi}=\alpha_{\Psi} \openone-\hat{C}_{\Psi},
\end{equation}
where $\alpha_{\Psi}=36.5$.

We use the method utilized for $\mathcal{W}_{\Phi}(\theta,\phi)$ to prove $\mathcal{W}_{\Psi}$ is a EW. First, we seek the witness operator $\mathcal{W}
_{\Psi}^{\text{p}}$. Through Ref. \cite{bouren}, the operator is given by: 
\begin{equation}
\mathcal{W}_{\Psi}^{\text{p}}=\frac{3}{4}\openone-\left|\Psi\right\rangle
\left\langle\Psi\right|.
\end{equation}
Then, we have to show that if a state $\rho$ satisfies $\text{Tr}(\mathcal{W}
_{\Psi}\rho)<0$, it also satisfies $\text{Tr}(\mathcal{W}_{\Psi}^{\text{p}
}\rho)<0$. We find that $\gamma_{\Psi}=30$ is such that $\mathcal{W}
_{\Psi}-\gamma_{\Psi}\mathcal{W}_{\Psi}^{\text{p}}\geq0$.

The sets of correlator operators $\hat{C}^{(x)}_{\Psi}$, $\hat{C}
^{(y)}_{\Psi}$, and $\hat{C} ^{(z)}_{\Psi}$ note that only three LMSs are used
in the witness operator $\mathcal{W}_{\Psi}$. The number of LMSs is smaller
than the required one, 15 LMSs, in Ref. \cite{bouren}. Moreover, the
robustness of the witness $\mathcal{W}_{\Psi}$ is specified by $\delta_{
\text{noise},\Psi}=15/88\simeq{0.170455}$. This result satisfies the
experimental requirement of robustness in Ref. \cite{bouren}.

\subsection{Detection of genuine multipartite entanglement for a four-level
tripartite system}

In order to show further utilities of the proposed approach, we proceed to
provide a witness to detect genuine multipartite entanglement close to a
four-level tripartite GHZ state \cite{4x3}: 
\begin{equation}
\left| \text{GHZ}_{4\text{x}3}\right\rangle =\frac{1}{2}\sum_{l=0}^{3}\left|
l\right\rangle _{1z}\otimes \left| l\right\rangle _{2z}\otimes \left|
l\right\rangle _{3z}.
\end{equation}
First of all, with the knowledge of the wave function represented in the
eigenbasis: $\left| l\right\rangle _{nz}$ for $n=1,2,3$, we have nine sets of
correlator operators for identifying quantum correlation between the $n^{
\text{th}}$ party and others, and are given by: 
\begin{equation}
\hat{C}_{nk,j}^{(z)}=(\hat{k}-\hat{s}_{kj})_{nz}\hat{k}_{pz}\hat{k} _{qz},
\end{equation}
for $j=1,...,9$; k=0,...,3; $n,p,q=1,2,3$, and $n\neq p\neq q$; where $\hat{s
} _{kj}=\hat{0},...,\hat{3}$; $\hat{k}\neq \hat{s}_{kj}$ and $\hat{s}
_{kj}\neq \hat{s}_{k^{\prime }j}$ for $k\neq k^{\prime }$; and $\hat{C}
_{nk,j}^{(z)}\neq \hat{C}_{nk,j^{\prime }}^{(z)}$ for $j\neq j^{\prime }$.
To show $\hat{C}_{nk,j}^{(z)}$ explicitly, let us take the following set of
operators numbered by $j=1$ for example: 
\begin{eqnarray}
\hat{C}_{n0,1}^{(z)}&=&(\hat{0}_{nz}-\hat{1}_{nz})\hat{0}_{pz}\hat{0}_{qz}, 
\nonumber \\
\hat{C}_{n1,1}^{(z)}&=&(\hat{1}_{nz}-\hat{2}_{nz})\hat{1}_{pz}\hat{1}_{qz}, 
\nonumber \\
\hat{C}_{n2,1}^{(z)}&=&(\hat{2}_{nz}-\hat{3}_{nz})\hat{2}_{pz}\hat{2}_{qz}, 
\nonumber \\
\hat{C}_{n3,1}^{(z)}&=&(\hat{3}_{nz}-\hat{0}_{nz})\hat{3}_{pz}\hat{3}_{qz}. 
\nonumber
\end{eqnarray}
Another example for the second set of operators, $j=2$, could be the
following one: 
\begin{eqnarray}
\hat{C}_{n0,2}^{(z)}&=&(\hat{0}_{nz}-\hat{2}_{nz})\hat{0}_{pz}\hat{0}_{qz}, 
\nonumber \\
\hat{C}_{n1,2}^{(z)}&=&(\hat{1}_{nz}-\hat{3}_{nz})\hat{1}_{pz}\hat{1}_{qz}, 
\nonumber \\
\hat{C}_{n2,2}^{(z)}&=&(\hat{2}_{nz}-\hat{0}_{nz})\hat{2}_{pz}\hat{2}_{qz}, 
\nonumber \\
\hat{C}_{n3,2}^{(z)}&=&(\hat{3}_{nz}-\hat{1}_{nz})\hat{3}_{pz}\hat{3}_{qz}. 
\nonumber
\end{eqnarray}
\newline
We progress to a correlation condition for the pure four-level tripartite GHZ
state by the following proposition:

\textbf{Proposition 2.} If the expectation values of $\hat{C}_{nk,j}^{(z)}$
for some state are all positive for $k=1,...,3$ under some $j$, the outcomes
of measurements for the party $n$ and the rest of the systems are correlated.

\textit{Proof.} If the $n^{\text{th}}$ party is independent of the rest of
the system, we can cast the expectation values of the operators $\hat{C}
_{nk,j}^{(z)}$ as 
\begin{eqnarray}
C_{nk,j}^{(z)}=(P(v_{n}=k)-P(v_{n}=s_{kj}))P(v_{p}=k,v_{q}=k).  \nonumber
\end{eqnarray}
Since $P(v_{p}=k,v_{q}=k)\geq 0$, $C_{nk,j}^{(z)}$ should not be all
positive. Thus $\hat{C}_{nk,j}^{(z)}>0$ for all $k$'s implies that the
measured outcomes for the party $n$ and the rest are correlated. Q.E.D.

All of the expectation values of the operators $\hat{C}_{nk,j}^{(z)}$
for the pure four-level tripartite GHZ state are given by $C_{nk,j,\text{GHZ}
_{4\text{x}3}}^{(z)}=1/4$, which are greater than zero. We then consider
that $\hat{C}_{nk,j}^{(z)}>0$ as a necessary condition of the state.

Secondly, if an observable with the eigenvector: 
\begin{equation}
\left| g\right\rangle_{nf}=\frac{1}{2}\sum_{h=0}^{3}\exp[-i\frac{2\pi h}{4}g]
\left| h\right\rangle_{nz},
\end{equation}
for $g=0,...,3$, is measured for each party $n=1,2,3$, we give nine sets of
correlator operators to identify quantum correlation between the $n^{\text{th
}}$ party and others: 
\begin{equation}
\hat{C} ^{(f)}_{nk,j}=(\hat{k}-\hat{s}_{kj})_{nf}\hat{V}_{klr},
\end{equation}
where 
\begin{equation}
\hat{V}_{klr}=\sum^{3}_{l,r=0}\delta[(k+l+r)\text{mod}\:4,0]\hat{l}_{pf}\hat{
r}_{qf},
\end{equation}
and definitions of $\hat{k}$, $\hat{s}_{kj}$, $n$, $p$, $q$, and $j$ are
same as the ones mentioned for $\hat{C}^{(z)}_{nk,j}$. For $j=1$, the set of
operators specified by the above equations could be: 
\begin{eqnarray}
\hat{C}_{n0,1}^{(f)}&=&(\hat{0}-\hat{1})(\hat{0}\hat{0}+\hat{1}\hat{3}+\hat{2
}\hat{2}+\hat{3}\hat{1}),  \nonumber \\
\hat{C}_{n1,1}^{(f)}&=&(\hat{1}-\hat{2})(\hat{0}\hat{3}+\hat{1}\hat{2}+\hat{2
}\hat{1}+\hat{3}\hat{0}),  \nonumber \\
\hat{C}_{n2,1}^{(f)}&=&(\hat{2}-\hat{3})(\hat{0}\hat{2}+\hat{1}\hat{1}+\hat{2
}\hat{0}+\hat{3}\hat{3}),  \nonumber \\
\hat{C}_{n3,1}^{(f)}&=&(\hat{3}-\hat{0})(\hat{0}\hat{1}+\hat{1}\hat{0}+\hat{2
}\hat{3}+\hat{3}\hat{2}).  \nonumber
\end{eqnarray}
For $j=2$, we could give the set of operators as follows: 
\begin{eqnarray}
\hat{C}_{n0,2}^{(f)}&=&(\hat{0}-\hat{2})(\hat{0}\hat{0}+\hat{1}\hat{3}+\hat{2
}\hat{2}+\hat{3}\hat{1}),  \nonumber \\
\hat{C}_{n1,2}^{(f)}&=&(\hat{1}-\hat{3})(\hat{0}\hat{3}+\hat{1}\hat{2}+\hat{2
}\hat{1}+\hat{3}\hat{0}),  \nonumber \\
\hat{C}_{n2,2}^{(f)}&=&(\hat{2}-\hat{0})(\hat{0}\hat{2}+\hat{1}\hat{1}+\hat{2
}\hat{0}+\hat{3}\hat{3}),  \nonumber \\
\hat{C}_{n3,2}^{(f)}&=&(\hat{3}-\hat{1})(\hat{0}\hat{1}+\hat{1}\hat{0}+\hat{2
}\hat{3}+\hat{3}\hat{2}).  \nonumber
\end{eqnarray}
Please note that, in order to have compact forms, we have omitted the
subscripts: $nf$, $pf$, and $qf$, from the above examples. A correlation
condition similar to the one discussed in proposition 2 is proposed by the
statement: if the expectation values of $\hat{C}_{nk,j}^{(f)}$ are all
positive for $k=1,...,3$ under some $j$, there are correlations between the
measured outcomes for the party $n$ and the rest of the systems. Since all
of the expectation values of the operators $\hat{C}_{nk,j}^{(f)}$ for the
pure four-level tripartite GHZ state are greater than zero, i.e., $C_{nk,j,
\text{GHZ}_{4\text{x}3}}^{(f)}=1/4$, the correlation condition: $\hat{C}
_{nk,j}^{(f)}>0$ is then necessary for the state.

Therefore, through a linear combination of all of the correlator operators
proposed above 
\begin{equation}
\hat{C}_{\text{GHZ}_{4\text{x}3}}=\sum^{3}_{n=1}\sum^{9}_{j=1}
\sum_{k=0}^{3}(1.5\hat{C}^{(z)}_{nk,j}+ \hat{C}^{(f)}_{nk,j}),
\end{equation}
the following witness operator detects genuine multipartite entanglement for
states close to $\left| \text{GHZ}_{4\text{x}3}\right\rangle$: 
\begin{equation}
\mathcal{W}_{\text{GHZ}_{4\text{x}3}}=\alpha_{\text{GHZ}_{4\text{x}3}} 
\openone-\hat{C}_{\text{GHZ}_{4\text{x}3}},
\end{equation}
where $\alpha_{\text{GHZ}_{4\text{x}3}}=40.5$.

We take an approach similar to the ones used in the previous proofs for EWs to
prove that the above witness operator detects genuine multipartite entanglement.
In order to show that if an experimental output state $\rho$ satisfies $
\text{Tr}(\mathcal{W}_{\text{GHZ}_{4\text{x}3}}\rho)<0$, the state $\rho$
also satisfies $\text{Tr}(\mathcal{W}_{\text{GHZ}_{4\text{x}3}}^{\text{p}
}\rho)<0$, firstly, we deduce that 
\begin{equation}
\mathcal{W}_{\text{GHZ}_{4\text{x}3}}^{\text{p}}=\frac{1}{4}\openone-\left|
\text{GHZ}_{4\text{x}3}\right\rangle\left\langle\text{GHZ}_{4\text{x}
3}\right|,
\end{equation}
by the method proposed in Ref. \cite{bouren}. Further, through the relation: 
$\mathcal{W}_{\text{GHZ}_{4\text{x}3}}-36\mathcal{W}_{\text{GHZ}_{4\text{x}
3}}^{\text{p}}\geq0$ for the proposed witness operator, we then conclude
that $\mathcal{W}_{\text{GHZ}_{4\text{x}3}}$ can be used to detect truly
multipartite entanglement.

Furthermore, when a state mixes with white noise the proposed EW is very
robust, and it detects genuine multi-partite entanglement if $p_{\text{noise}
}<0.4$. Thus, two local measurement settings are sufficient to detect
genuine four-level tripartite entanglement around a pure four-level
tripartite GHZ state.

\section{BI for arbitrary high-dimensional bipartite systems}

In order to derive a new BI, we will begin with specifications of
correlation conditions for quantum correlation of a two-qudit entangled
state. Then, we will proceed to verify that any local hidden variable theory
cannot reproduce the correlations embedded in the entangled state. This
approach is novel and opposite to the one in Ref. \cite
{collins}.

First, to specify the quantum correlation embedded in the two-qudit
entangled state, 
\begin{equation}
\left| \psi _{d}\right\rangle =\frac{1}{\sqrt{d}} \sum_{l=0}^{d-1}\left|
l\right\rangle_{1z}\otimes\left|l\right\rangle_{2z},
\end{equation}
we describe the wave function in the following eigenbasis of some observable 
$\hat{V}_{k}^{(q)}$: 
\begin{equation}
\left|l\right\rangle_{kq}=\frac{1}{ \sqrt{d}}\sum^{d-1}_{m=0}\exp[i\frac{
2\pi m}{d}(l+n_{k}^{(q)})]\left| m\right\rangle_{kz},
\end{equation}
for $k,q=1,2$, where $n_{1}^{(1)}=0$, $n_{2}^{(1)}=1/4$, $n_{1}^{(2)}=1/2$,
and $n_{2}^{(2)}=-1/4$ correspond to four different LMSs, $M_{ij}=(\hat{V}
_{1}^{(i)},\hat{V} _{2}^{(j)})$ for $i,j=1,2$. From our knowledge of the
four different representations of the state $\left|\psi_{d}\right\rangle$,
we give four sets of correlators of quantum correlation: 
\begin{eqnarray}
&&C_{m}^{(12)}=P(v_{1}^{(1)}=(-m)\,\text{mod}\:d,v_{2}^{(2)}=m)  \nonumber \\
&&\quad\quad\quad-P(v_{1}^{(1)}=(1-m)\text{mod}\:d,v_{2}^{(2)}=m), \\
&&C_{m}^{(21)}=P(v_{1}^{(2)}=(d-m-1)\text{mod}\:d,v_{2}^{(1)}=m)  \nonumber
\\
&&\quad\quad\quad-P(v_{1}^{(2)}=(-m)\text{mod}\:d,v_{2}^{(1)}=m), \\
&&C_{m}^{(qq)}=P(v_{1}^{(q)}=(-m)\,\text{mod}\:d,v_{2}^{(q)}=m)  \nonumber \\
&&\quad\quad\quad-P(v_{1}^{(q)}=(d-m-1)\text{mod}\:d,v_{2}^{(q)}=m),\ \ \ 
\end{eqnarray}
for $m=0,1,...,d-1$ and $q=1,2$. The superscripts, $(ij)$, $(i)$, and $(j)$,
indicate that some LMS, $M_{ij}$, has been selected. For the pure state $
\left|\psi _{d}\right\rangle$ under $M_{ij}$, the correlator $
C_{m}^{(ij)}$ can be evaluated analytically \cite{collins} and are given by 
\begin{equation}
C_{m,\psi_{d}}^{(ij)}=\frac{1}{2d^{3}}(\csc ^{2}(\pi /4d)-\csc ^{2}(3\pi
/4d)),
\end{equation}
where $\csc(h)$ is the cosecant of $h$. Since $C_{m,\psi_{d}}^{(ij)}>0$ for
all $m$'s with any finite value of $d$, we ensure that there are
correlations between outcomes of measurements performed on the state $
\left|\psi _{d}\right\rangle $ under four different LMSs. The proof of this
statement is similar to that for proposition 2. Hence the correlation
conditions: 
\begin{equation}
C_{m}^{(ij)}>0,
\end{equation}
are necessary for the pure two-qudit entangled state $\left|\psi
_{d}\right\rangle$.

Thus, we take the summation of all $C_{m}^{(ij)} $ 's, 
\begin{equation}
C_{d}=C^{(11)}+C^{(12)}+C^{(21)}+C^{(22)},
\end{equation}
where $C^{(ij)}=\sum_{m=0}^{d-1}C_{m}^{(ij)}$, as an identification of the
state $\left|\psi_{d}\right\rangle$. We can evaluate the summation of all $
C_{m}^{(ij)}$'s for the state $\left|\psi _{d}\right\rangle$, and then we
have 
\begin{equation}
C_{d,\psi _{d}}=\frac{2}{d^{2}}(\csc ^{2}(\pi /4d)-\csc ^{2}(3\pi /4d)).
\end{equation}
One can find that $C_{d,\psi _{d}}$ is an increasing function of $d$. For
instance, if $d=3$, one has $C_{3,\psi _{3}}\simeq 2.87293$. In the limit of
large $d$, we obtain, $\lim_{d\rightarrow \infty }C_{d,\psi _{d}}=(16/3\pi
)^{2}\simeq 2.88202$.

We proceed to consider the maximum value of $C_{d}$ for local hidden
variable theories which is denoted by $C_{d,\text{LHV}}$. The following
proof is based on deterministic local models which are specified by \emph{
fixing} the outcome of all measurements. This consideration is general since
any probabilistic model can be converted into a deterministic one \cite
{percival}. Substituting a \emph{fixed} set, $( \tilde{v}_{1}^{(1)},\tilde{v}
_{2}^{(1)},\tilde{v}_{1}^{(2)},\tilde{v} _{2}^{(2)})$, into 
\begin{eqnarray}
C^{(ij)}_{m}=P(v_{1}^{(i)}=\alpha_{m}^{(ij)},v_{2}^{(2)}=m)-P(v_{1}^{(i)}=
\beta_{m}^{(ij)},v_{2}^{(j)}=m),  \nonumber
\end{eqnarray}
where $\alpha_{m}^{(ij)}$ and $\beta_{m}^{(ij)}$ denote the values involved
in the Eqs. (50), (51), and (52), then we have the result: 
\begin{equation}
C^{(ij)}_{m,\text{LHV}}=\delta[\alpha_{m}^{(ij)},\tilde{v}_{1}^{(i)}]\delta[
m,\tilde{v}_{2}^{(j)}]-\delta[\beta_{m}^{(ij)},\tilde{v}_{1}^{(i)}]\delta[m,
\tilde{v}_{2}^{(j)}],
\end{equation}
where $\delta \lbrack x,y]$ denotes the Kronecker delta symbol. Accordingly, 
$C_{d}$ for local hidden variable theories turns into 
\begin{eqnarray}
&&C_{d,\text{LHV}}  \nonumber \\
&&=\delta\lbrack (\tilde{v}_{1}^{(1)}+\tilde{v}_{2}^{(1)})\text{mod}
\:d,0]-\delta \lbrack -(\tilde{v}_{1}^{(1)}+\tilde{v}_{2}^{(1)})\text{mod}
\:d,1]  \nonumber \\
&&+\delta\lbrack (\tilde{v}_{1}^{(1)}+\tilde{v}_{2}^{(2)})\text{mod}
\:d,0]-\delta \lbrack (\tilde{v}_{1}^{(1)}+\tilde{v}_{2}^{(2)})\text{mod}
\:d,1]  \nonumber \\
&&+\delta\lbrack (\tilde{v}_{1}^{(2)}+\tilde{v}_{2}^{(2)})\text{mod}
\:d,0]-\delta \lbrack -(\tilde{v}_{1}^{(2)}+\tilde{v}_{2}^{(2)})\text{mod}
\:d,1]  \nonumber \\
&&+\delta\lbrack -(\tilde{v}_{1}^{(2)}+\tilde{v}_{2}^{(1)})\text{mod}
\:d,1]-\delta \lbrack (\tilde{v}_{1}^{(2)}+\tilde{v}_{2}^{(1)})\text{mod}
\:d,0].  \nonumber \\
\end{eqnarray}
There are three non-vanishing terms at most among the four positive delta
functions, and there exist four cases for it, for example, one is that if $
\delta \lbrack ( \tilde{v}_{1}^{(1)}+\tilde{v}_{2}^{(1)})\text{mod}
\:d,0]=\delta \lbrack ( \tilde{v}_{1}^{(1)}+\tilde{v}_{2}^{(2)})\text{mod}
\:d,0]=\delta \lbrack ( \tilde{v}_{1}^{(2)}+\tilde{v}_{2}^{(2)})\text{mod}
\:d,0]=1$ is assigned, we obtain $\tilde{v}_{2}^{(1)}=\tilde{v}_{2}^{(2)}$
and then deduce that $\delta \lbrack -(\tilde{v}_{1}^{(2)}+\tilde{v}
_{2}^{(1)})\text{mod}\:d,1]=0$ . We also know that there must exist one
non-vanishing negative delta function and three vanishing negative ones in
the $C_{d,\text{LHV}}$ under the same condition. In the example, the case is 
$\delta \lbrack (\tilde{v} _{1}^{(2)}+\tilde{v}_{2}^{(1)})\text{mod}\:d,0]=1$. With these facts, we conclude that $C_{d,\text{LHV}}\leq 2$. One can check
other three cases for the four positive $\delta$ functions, and then they
always result in the same bound. Thus, we realize that $C_{d,\psi _{d}}>C_{d,
\text{LHV} }$ and the quantum correlations are stronger than the ones
predicted by the local hidden variable theories

For $d=2$, the proposed inequality $C_{2,\text{LHV}}\leq 2$ can be
expressed explicitly in the form
\begin{equation}
\tilde{\text{C}}^{(11)}+\tilde{\text{C}}^{(12)}+\tilde{\text{C}}^{(22)}-
\tilde{\text{C}}^{(21)}\leq 2,
\end{equation}
where $\tilde{\text{C}}^{(ij)}=\sum_{k=0}^{1}(-1)^{k}\delta \lbrack (\tilde{v
}_{1}^{(i)}+\tilde{v}_{2}^{(j)})\text{mod}\:d,k]$, and then we obtain the
result which is known as the \textit{CHSH inequality} after the discovery of
Clauser, Horne, Shimony, and Holt \cite{chsh}. On the other hand, from the
quantum mechanical point of view, we have a violation of the CHSH inequality
by $C_{2,\psi _{2}}=2\sqrt{2}$.

A surprising feature of the new inequality is that the total number of
detection events required for analyses by each of the presented correlation
functions $C^{(ij)}$ is only $2d$, which is much smaller than the result, $
O( d^{2})$, shown in Ref. \cite{fu}. This implies that the proposed
correlation functions contain only the dominant terms to identify
correlations. However, the proposed BI is \textit{non-tight} from a
geometric point of view \cite{tight}. Since the number of linear independent
generators contained in the hyperplane $C_{d,\text{LHV}}=2$ is only $4d$ 
\cite{further} which is smaller than $4d(d-1)$ involved in the condition of 
\textit{tightness} \cite{tight}, the BI is non-tight.

Furthermore, if an experimental output state suffered from white noise and
turned into a mixed one with the form 
\begin{eqnarray}
\rho=\frac{p_{\text{noise}}}{d^{2}}\openone +(1-p_{\text{noise}
})\left|\psi_{d}\rangle\langle\psi_{d}\right|,  \nonumber
\end{eqnarray}
\newline
the value of $C_{d}$ for the state $\rho$ becomes $C_{d,\rho}=(1-p_{\text{
noise}})C_{d,\psi_{d}}$. If the criterion, $C_{d,\rho}>2$, i.e., 
\begin{equation}
p_{\text{noise}}<1-\frac{2}{C_{d,\psi_{d}}},
\end{equation}
is imposed on the system, one ensures that the mixed state still exhibits
quantum correlations in outcomes of measurements. For instance, to maintain
the quantum correlation for the limit of large $d$, the system must have $p_{
\text{noise}}<0.30604$.

On the other hand, it is worth comparing the noise tolerance of $C_{d}$
with the one of the following EW: 
\begin{equation}
\mathcal{W}_{\psi_{d}}^{\text{p}}=\frac{1}{d}\openone-\left|\psi_{d}\right
\rangle\left\langle\psi_{d}\right|.
\end{equation}
Let the noise fraction be the form: $p_{\text{noise}}=1-\epsilon$, where $
\epsilon$ is a positive parameter. Then satisfying the condition of
entanglement $\text{Tr}(\mathcal{W}^{\text{p}}_{\psi_{d}}\rho)<0$ implies
that $\epsilon>1/(d+1)$. Therefore, in the case where $d\rightarrow\infty$,
any state with $p_{\text{noise}}<1$ is detected as an entangled one. Hence,
there is a significant difference between the noise tolerance of $C_{d}$ and
the one of $\mathcal{W}_{\psi_{d}}^{\text{p}}$ in the limit of large $d$.

\section{Summary}

Through the novel necessary condition of quantum correlation we develop a
systematic approach to derive correlator operators for BIs and EWs. The new $d
$-level bipartite BI is strongly resistant to noise and can be tested with
fewer analyses of measurement outcomes. The proposed EWs for the generalized
GHZ, four-qubit singlet, and four-level tripartite GHZ states are robust to
noise and require fewer experimental efforts to be realized. Therefore, the
correlation conditions for quantum correlation involved in the approach to
construct correlator operators utilized in EWs and BIs can be considered as a
connection between them. The generality of the approach widely cover several
(different) tasks of entanglement detections and pave the way for further
studies on entangled qudits.

We thank J.W. Pan, Z.B. Chen, and Q. Zhang for useful discussions. This work
is supported partially by the National Science council of Taiwan under the
grand number NSC95-2119-M-009-030.

\end{document}